\documentclass{emulateapj}
\usepackage{apjfonts}
\usepackage{epsfig}

\def\msunh{\mbox{$h^{-1}\ M_\odot$}}

\shorttitle{Theoretical predictions of DM halos}
\shortauthors{BETANCORT-RIJO ET AL.}

\begin{document}

\title{Detailed theoretical predictions for the outskirts of dark matter halos}

\author{Juan E. Betancort-Rijo$^{1,2}$, Miguel A. Sanchez-Conde$^3$, Francisco Prada$^3$ and Santiago G. Patiri$^1$}

\affil{$^{1}$ Instituto de Astrof\'{i}sica de Canarias, V\'{i}a L\'{a}ctea s/n, 
 La Laguna, Tenerife, E38200, Spain}
\affil{$^{2}$ Facultad de Fisica, Universidad de La Laguna, Astrofisico Francisco Sanchez, s/n, La Laguna Tenerife, E38200, Spain}
\affil{$^{3}$ Instituto de Astrof\'{i}sica de Andaluc\'ia (CSIC), E18008, Granada, Spain}

\email{jbetanco@iac.es}

\begin{abstract}

In the present work we describe the formalism necessary to derive the properties of dark matter halos beyond two
virial radius using the spherical collapse model (without shell crossing), and provide the framework for the 
theoretical prediction presented in Prada et al.~(2005). We show in detail how to obtain within this 
model the probability distribution for the spherically-averaged enclosed density at any radii $P(\delta,r)$. Using this probability
distribution, we compute the most probable and mean density profiles, which turns out to differ considerably from each 
other. We also show how to obtain the typical profile, as well as the probability distribution and mean profile for the 
spherically averaged radial velocity. Three probability distributions are obtained: a first one is derived using a 
simple assumption, that is, if $Q$ is the virial radius in Lagrangian coordinates, then the enclosed linear contrast 
$\delta_{l}(q)$ must satisfy the condition that $\delta_{l}(q=Q)=\delta_{vir},$ where $\delta_{vir}$ is the linear 
density contrast within the virial radius $R_{vir}$ at the moment of virialization. Then we introduce an additional 
constraint to obtain a more accurate $P(\delta,r)$ which reproduces to a higher degree of precision 
the distribution of the spherically averaged enclosed 
density found in the simulations. This new constraint is that, for a given $q>Q$, $\delta_l(q) < \delta_{vir}$. 
A third probability distribution, the most accurate, is obtained imposing the strongest constraint that 
$\delta_{l}(q) < \delta_{vir} \quad \forall~ q > Q$, which means that there are no radii larger than $R_{vir}$ where 
the density contrast is larger than that used to define the virial radius. Finally, we compare our theoretical 
predictions for the mean density and mean velocity profiles with the results found in the simulations.
\end{abstract}


\section{Introduction}

The study of the density profile of cold dark matter halos beyond the virial radius 
is a subject of considerable relevance. From an observational point of view, knowledge 
of the shape of the density profile far beyond the virial radius is essential for an 
appropriate interpretation of gravitational lensing phenomena (e.g. Smith et al.~2001; 
Guzik \& Seljak 2002; Hoekstra et al.~2004; Sheldon et al.~2004), the pattern of Lyman alpha absorption around 
virialized systems (e.g. Barkana 2004; Bajtlik, Duncan \& Ostriker 1988) as well as 
the motion of satellite galaxies as a test for dark matter distribution at large radii 
(Zaritsky \& White 1994, Zaritsky et al.~1997; Prada et al.~2003, Brainerd 2004; Conroy et al.~2004).
 From the theoretical point of view, the study of the properties of dark matter halos at several 
virial radius in cosmological simulations provides an excellent benchmark for developing and testing the basic 
theoretical framework which will be decisive for a full understanding of the physical origin and formation of the 
$\Lambda$CDM halos.

Understanding halo properties involves a set of theoretical 
considerations. First, we have the issue of choosing the correct initial 
density profile. Also, there is the question of which processes are
relevant to the gravitational evolution of the initial profile: is 
only the spherical collapse what matters or is triaxiality
important? up to which radius can we use the standard spherical collapse 
model without shell crossing? are highly asymmetrical processes, 
like merging, relevant? In order to answer these questions it is very convenient 
to focus first on those properties of the halos which involve the fewest theoretical uncertainties. 

The dark matter density profiles at several virial radius 
are particularly suitable to check whether the spherical collapse model can provide 
accurate predictions (see Prada et al.~2005). In fact, it has been shown that the spherical collapse model 
reproduces very well the relationship between the small values of the spherically-averaged 
enclosed density at those large distances and the radial velocity 
\citep{lilje}. 

We define the spherically-averaged enclosed density as:
\begin{displaymath}
\frac{\rho(<r)}{<\rho_{m}>} = 1+\delta
\end{displaymath}
where $\delta$ is the enclosed density contrast and $<\rho_{m}>$ the average matter density in the Universe.
We can also define the spherically-averaged local density as:
\begin{displaymath}
\frac{\rho(r)}{<\rho_{m}>} = 1+\delta'
\end{displaymath}
where $\delta'$ is the density contrast in a narrow shell of radius r. We can then obtain the 
density contrast $\delta'$ from the enclosed density contrast $\delta$ using the relation:
\begin{displaymath}
\delta'(r) = \frac{1}{3r^{2}}~\frac{d}{dr}(r^{3}\delta(r)).    
\end{displaymath} 

Despite to all the effort done to understand the central dense regions of the dark matter halos in cosmological simulations,
not much attention has been devoted to the study of the regions beyond the formal virial radius, i.e. the radius 
within which the spherically-averaged enclosed density is equal to some specific value.  
The main goal of the work presented in this paper is focused on the 
outskirts of the dark matter halos, where the correct evolution of the spherically-averaged enclosed 
density profiles can essentially be obtained using the standard (without shell crossing) 
spherical collapse model. This model, first developed by Gunn \& Gott (1972) and Gunn (1977), 
describes the collision-less collapse of a spherical perturbation in an expanding background. They introduced 
for the first time the cosmological expansion and the role of adiabatic invariance in the formation 
of individual objects. Later, Fillmore \& Goldreich (1984) found analytical predictions for the density of collapsed 
objects seeded by scale-free primordial perturbations in a flat universe. Hoffman \& Shaham (1985) generalized 
these solutions to realistic initial conditions in flat and open Friedmann models. Some studies have been done 
to include more realistic dynamics of the growth process (e.g. Padmanabhan 1996; Avila-Reese, Firmani \& Hern\'andez 
1998; Lokas 2000; Subramanian, Cen \& Ostriker 2000). 

There are plenty of works in the literature using the spherical collapse model to predict
the density profiles of dark matter halos mainly focused on explaining their central regions. 
For example Bertschinger (1985) used the spherical collapse with shell crossing 
to obtain the density profiles resulting from initial power law density profiles. 
Lokas \& Hoffman (2000) considered more general initial profiles. The effect of non-radial 
motions has also been widely treated (see Ryden \& Gunn 1987; Gurevich \& Zybin 1988; 
Avila-Reese et al.~1998; White \& Zaritsky 1992; Sikivie et al.~1997; 
Nusser 2001; Hiotelis 2001). Some of these authors have used arbitrary
initial profiles, while others have assumed the mean initial profile around 
density maxima (Bardeen et al.~1986, BBKS). In all these works angular momentum is 
introduced by hand, although more recently it has been done in a more natural way 
(Nusser 2001; Ascasibar et al.~2004). Concerning to the outer parts of the dark matter halos, 
only Barkana (2004) has adopted an appropriate initial profile, but only for a restricted type of density profile (the typical profile). 
The more recent work by Prada et al.~(2005) have obtained predictions for the mean and most probable density 
profiles and have provided a detailed comparison with cosmological simulations.

A proper understanding of the physics of dark matter halos involves predicting
correctly not only the mean halo density profile for any given mass 
but also the whole probability distribution for the enclosed
density contrast at any given radii, $P(\delta,r)$. A first attempt to determine it can be found
 in Prada et al.~(2005), where it has been shown to be generally in good agreement with the cosmological simulations.
Nevertheless, this probability distribution shows, at any radius, a longer tail for large values of $\delta$, as compared to that from 
simulations at any radius. In this paper we present a more accurate prediction for the probability distribution 
$P(\delta,r)$ that constitutes the main new result of this work. We also give in detail the theoretical 
background of the predictions presented in Prada et al.~(2005). The agreement of our new predictions with 
the simulations is excellent even in the tail of the distribution. 
Furthermore, we also compute the radial velocity probability distribution and the mean radial velocity profile.

The work is organized as follows. In section 2 we present our theoretical framework and obtain the typical 
density profile of dark matter halos. In section 3 we show in detail how to obtain the probability distribution 
for the spherically-averaged enclosed density contrast at a given radii, $P(\delta,r)$, presented in 
Prada et al.~(2005). The most probable and mean profiles are derived. In section 4, we compute the 
probability distribution and mean profile for the spherically 
averaged radial velocity. In section 5, we obtain more accurate probability distributions
than that used in the previous sections, and compare again with that found in the simulations. Final remarks are given  
in section 6.

\section{The typical density profile of Dark Matter Halos}\label{Sec1}

The present spherically-averaged enclosed density profiles attains a density contrast value of $\Delta_{vir}$ at  
certain radius, 
the so called virial radius. At larger radii the density contrast must be, by definition, smaller than $\Delta_{vir}$,  
otherwise
the virial radius would be larger than its nominal value.

We shall now make some comments on the values of $\Delta_{vir}$ and $\delta_{vir}$ that we use:
although at several virial radius the spherically-averaged enclosed linear and actual densities 
are related by the spherical collapse model, the same does not apply within the virial radius. 
The spherically-averaged enclosed density contrast within one virial radius, $\Delta_{vir}$, and
the corresponding enclosed linear density contrast, $\delta_{vir}$, are
not related as homologous quantities at larger radii, because at one virial radius 
shell crossing has already becomes important. Consequently, the value of $\delta_{vir}$ corresponding to 
$\Delta_{vir}$=340 (the value we adopted to define the virial radius) is somewhat uncertain. 
As a result of work still in progress we will be able to provide 
the precise values for $\delta_{vir}$ and determine its possible small dependence 
on mass. Here we use for all masses $\delta_{vir}$=1.9, a value that leads to very good results and
that may be inferred from the fact that when $\Delta_{vir}$=180, $\delta_{vir}$ seems to be close to 1.68 for
all cosmologies \citep{jenkins}.

It must be noted that for all our predictions it is irrelevant whether the value of 
$\Delta_{vir}$ that we use actually corresponds to the
virial density contrast or not. By virial density contrast is usually meant 
the enclosed density contrast within the largest radii so that we have statistical 
equilibrium. The precise value of $\Delta_{vir}$ corresponding to this definition is still problematic but, to our purposes, 
it can be chosen freely to define a conventional "virial radius". Since we have used numerical simulations 
with $\Delta_{vir}$ equal to 340, we will take the same value by default in all the calculations.

Let $\delta_{l}(q,i)$ be a realization of the spherically-averaged initial enclosed density profile around a protohalo 
(with a given present virial radius, $R$) linearly extrapolated to the present, where $q$ is the lagrangian distance 
from the center of the halo to the given point and $i$ is an index running over realizations. Any realization of the 
initial profile may be transformed using the standard spherical collapse model (without shell-crossing). We can use the 
relationship between the linear value of the density contrast within a sphere, $\delta_{l}$, and the actual density 
contrast within that sphere, $\delta$, for the concordant cosmology (Sheth and Tormen (2002)):

\begin{equation}
\delta_{l}(\delta) = \frac{1.676}{1.68647}\bigg[ 1.68647 - \frac{1.35}{(1+\delta)^{2/3}} - 
\frac{1.12431}{(1+\delta)^{1/2}} + \frac{0.78785}{(1+\delta)^{0.58661}}\bigg]    \label{eq1}
\end{equation}

or, rather its inverse function $\delta(\delta_{l})$ (Patiri et al.(2004) expression (4)):

\begin{displaymath}
\delta(\delta _l)=0.993[(1-0.607(\delta _{l}-6.5\times 10^{-3}(1-\theta(\delta_{l})+  \nonumber
\end{displaymath}
\begin{equation}
+\theta(\delta_{l}-1.55))\delta_l^2))^{-1.66}-1]   \label{eq2}
\end{equation}
being $\theta$ the step function:
\begin{displaymath}
 \theta(x)=\left \{ \begin{array}{ll}
  1 & \rm{if~x>0}\\
  0 & \rm{if~x\leq 0}
 \end{array} \right.
\end{displaymath}

Transforming for every shell $\delta_{l}$ and $q$ into $\delta$ and $r$ (the Eulerian radius of the shell) we may  
obtain, 
in parametric form, the initial profile $\delta_{l}(q,i)$ spherically evolved, $\delta(r,i)$.

\begin{equation}
\delta(r,i)=\delta(\delta_{l}(q,i)) ; \qquad r=[1+\delta(\delta_{l}(q,i))]^{-1/3}q    \label{eq3}
\end{equation}

This two equations gives $\delta(r,i)$ implicitly with $q$ as parameter.

We must now eliminate all linear profiles leading to present profiles which attains an enclosed density contrast larger  
than (or equal to) $\Delta_{vir}$ at a radii larger than the nominal virial radius. The ensemble of remaining halos allow  
us to obtain the predictions of the standard spherical collapse model for any statistics. For example, we shall obtain the  
predictions for the mean profile:

\begin{displaymath}
\bar{\delta}(r) \equiv <\delta(r,j)>_{j}
\end{displaymath}

That is, the average over all remaining halos ($j$ runs over these halos). We shall also consider the most probable 
profile, $\delta_{p}(r)$, that is, the profile that associates with every value of $r$ the $\delta$ value with the  
largest probability density.

The procedure described above to obtain the spherical model predictions (i.e. realizations of the linear profile evolved with 
expression (\ref{eq3})) serve mostly to the purpose of clarifying the meaning of those predictions and 
we only use it as a test to the analytical expressions. In practise, we shall use another procedure to 
obtain directly $\overline\delta(r)$, $\delta_{p}(r)$ and, in fact, the whole probability distribution 
for the value of $\delta$ at a given value of r, $P(\delta,r)$.

Before dealing with the detailed predictions just mentioned we consider a simpler prediction which shall help 
clarifying the rest of the work, and which is related to previous approaches \citep{Barkana}.

Consider the ensemble of all halos, $\delta(r,j)$, attaining a $\delta$ value equal to $\Delta_{vir}$ at a virial radius and  
smaller 
values for larger radii. If we transform back these profile to their linear counterpart, we 
obtain the ensemble $\delta_{l}(q,j)$. Let us now take the average over this ensemble (now for a fixed value of $q$):

\begin{displaymath}
\bar{\delta}(q) \equiv <\delta(q,j)>_{j}
\end{displaymath}

Evolving this profile by means of the spherical collapse model we obtain a profile which we call typical profile and 
represent by $\delta_{t}(r)$, that is, this profile is simply the mean profile in the initial conditions spherically  
evolved.

Note that the typical density profile is defined because of its simplicity and not because it constitutes a 
prediction for any specific statistics of the actual halos. Nevertheless, it should not be very different 
from the most probable profile. Later we will study how these two profiles differ from each other. This profile 
definition is the same as that used by Barkana (2004), but we use a different approximation to derive it.

To obtain the mean linear density profile $\overline\delta_{l}(q)$ subject to the condition that in the present  
enclosed density 
profile the virialization density contrast, $\Delta_{vir}$, is not attained beyond the virial radius, 
Barkana used barrier penetration results. He obtained the probability distribution for $\delta_{l}$(q) 
given the conditions:

\begin{displaymath}
\delta_{l}(Q)=\delta_{vir} ; \qquad \delta_{l}(q)<\delta_{vir} \qquad \forall q>Q
\end{displaymath}
\begin{displaymath}
Q\equiv R_{vir}(1+\Delta_{vir})^{1/3}
\end{displaymath}

where $R_{vir}$ is the virial radius (Q the corresponding Lagrangian radius), and $\delta_{vir}$ is 
the linear counterpart of $\Delta_{vir}$. Then, by averaging over all $\delta_{l}$ values smaller 
than $\delta_{vir}$, he obtained the mean linear profile, $\overline\delta_{l}(q)$, with those 
conditions. However, he had to make some simplifications, the most relevant one is that he used a 
sharp filter in k-space, rather than the top hat filter in ordinary space which is the natural one in this context.

In our approach we first obtain the probability for $\delta_{l}(q)$ only with the condition  
$\delta_{l}(Q)=\delta_{vir}$:

\begin{equation}
P(\delta_{l},q) \equiv  P(\delta_{l}(q)|\delta_{l}(Q)=\delta_{vir})=\frac{ \exp \left( -\frac{1}{2} 
\frac{\big(\delta_{l}(q)- \frac{\sigma_{12}}{\sigma_{1}^2}\delta_{vir}\big)^2}{g}\right)}{\sqrt{2\pi}~g^{\frac{1}{2}}}  
\label{eq4}
\end{equation}

where

\begin{displaymath}
g(q) \equiv \left(\sigma_{2}^{2}-\frac{\sigma_{12}^2(q)}{\sigma_{1}^2}\right) 
\end{displaymath}
\begin{displaymath}
\sigma_{1}\equiv \sigma(Q) ; \qquad \sigma_{2}\equiv \sigma(q)
\end{displaymath}

\begin{displaymath}
(\sigma(x))^{2}=\frac{1}{2\pi^{2}}\int_{0}^{\infty}\mid\delta_{k}\mid^{2}W_{T}^2(xk)~k^{2}~dk
\end{displaymath}

\begin{displaymath}
\sigma_{12}=\sigma_{12}(q)=\frac{1}{2\pi^{2}}\int_{0}^{\infty}\mid\delta_{k}\mid^{2}W_{T}(qk)~W_{T}(Qk)~k^2~dk
\end{displaymath}

\begin{displaymath}
W_{T}(x)=\frac{3(sin~x~-x~cos~x~)}{x^3}
\end{displaymath}

where $|\delta_{k}|^{2}$ stands for the power spectra of the density fluctuations linearly extrapolated to the  
present.

It is convenient to use a simple and accurate approximation for $\sigma_{12}(q)$:

\begin{equation}
\frac{\sigma_{12}(q)}{(\sigma(Q))^{2}} \simeq e^{-b(Q)\big((\frac{q}{Q})^2-1\big)}  \label{eq5}
\end{equation}

where $b(Q)$ is a coefficient depending on the the size of the halo, $Q$:

\begin{displaymath}
b(Q)=-\frac{1}{2}~\frac{d\ln \sigma(x)}{d\ln x} \bigg|_{x=Q}
\end{displaymath}

If no restriction other than $\delta_{l}(Q)=\delta_{vir}$ were imposed on $\delta_{l}(q)$ the mean linear profile  
would be:

\begin{displaymath}
\overline\delta_{l}(q)=\int_{-\infty}^{\infty}P(\delta_{l}(q)/\delta_{l}(Q)=\delta_{vir})~\delta_{l}(q)~d(\delta_{l}(q))=\delta_{vir}\frac{\sigma_{12}(q)}{\sigma(Q)} \equiv \delta_{0}(q)
\end{displaymath}

However, we are interested on the mean profiles satisfying also $\delta_{l}(q)<\delta_{vir}$ for $q>Q$. So, we must  
use as mean profile:

\begin{eqnarray}
\overline\delta_{l}(q)=\frac{\int_{-\infty}^{\delta_{vir}}P(\delta_{l}(q)/\delta_{l}(Q)=\delta_{vir})~\delta_{l}(q)~d(\delta_{l}(q))}{\int_{-\infty}^{\delta_{vir}}P(\delta_{l}(q)/\delta_{l}(Q)=\delta_{vir})~d(\delta_{l}(q))} \label{eq6} \\ \nonumber \\
 = 
\delta_{0}(q)-\frac{\overline{\sigma}(q)~e^{-\frac{1}{2}~\frac{\big(\delta_{vir}-\delta_{0}(q)\big)^2}{(\overline{\sigma}(q))^2}}}{1-\frac{1}{2}erfc\big(\frac{\delta_{vir}-\delta_{0}(q)}{\sqrt{2}~\overline{\sigma}(q)}\big)}  \nonumber
\end{eqnarray}

\begin{displaymath}
\overline{\sigma}(q) \equiv (g(q))^{1/2}
\end{displaymath}

For arbitrarily massive halos $\sigma(Q)<<1$. So, since for the relevant $q$ values ($q > Q$) $\overline{\sigma}(q)<\sigma(Q)$, $\delta_{l}(q)$ is simply given 
by $\delta_{0}(q)$. However, for halos with $\sigma(Q) \gtrsim 1$, $\delta_{l}(q)$ is substantially steeper 
than $\delta_{0}(q)$, resulting in steeper present density profiles for smaller masses. This result has previously 
been advanced by Barkana \citep{Barkana} and have been confirmed by means of numerical simulations \citep{Paquin}.

It must be noted that the profile given by expression (\ref{eq6}) is not exactly the mean linear profile 
implicit in the definition of typical profile. Note that at each value of q, $\delta_{l}$ is constrained 
to lay below $\delta_{vir}$ but the probability distribution upon which this constraint is imposed does 
not account for the fact that the profile lies below $\delta_{vir}$ at any other value of q larger than Q. 
This is, however, a good approximation, because the most relevant part of the present density profile (to $\approx$ 
10 virial radius) corresponds to a narrow region in Lagrangian coordinates ($\approx1.5Q$). The value of 
$\delta_{l}$ for q between Q and 1.5Q are strongly correlated. So, if we impose the condition 
$\delta_{l}<\delta_{vir}$ at, for example, $q=1.25Q$, the probability that the same condition holds at any $q$ is  
close to one.

Once we have $\bar{\delta}_{l}(q)$ all we need to do is to evolve it with the spherical collapse model. Using 
equations (\ref{eq3}) we may write (Patiri et al.(2004) expression (20)):

\begin{equation}
\delta(r)=\delta(\overline\delta_{l}(q)) ; \qquad q \equiv r~[1+\delta(r)]^{\frac{1}{3}} \label{eq7}
\end{equation}

where the right hand side of this equation is simply the function defined in expression (\ref{eq2}) 
evaluated at $\overline\delta_{l}(q)$ (given by expression (\ref{eq6})). For each value of $r$ we 
must solve this equation for the variable $\delta(r)$. Applying the function defined in expression 
(\ref{eq1}), which is the inverse of that defined in (\ref{eq2}), to both sides of this equation we have:

\begin{equation}
\delta_{l}(\delta(r))=\overline\delta_{l}(q) ; \qquad q \equiv r~(1+\delta(r))^{\frac{1}{3}}   \label{eq8}
\end{equation}

where the left hand side is expression (\ref{eq1}) evaluated at $\delta(r)$. This equation is usually 
simpler to solve than equation (\ref{eq7}) and is the one we used in Prada et al.(2005). It must be noted, however, that, 
if one intends to generate the whole profile rather than to obtain $\delta(r)$ for some specific value of $r$, 
it is not necessary to solve equation (\ref{eq8}), since one may obtain all couples of values of $\delta(r)$, $r$ 
using (\ref{eq7}) and running over all values of $q$.
The profile obtained in this way is the typical enclosed density contrast profile, $\delta_{t}(r)$. 
To obtain the typical density contrast profile, $\delta'_{t}(r)$, we may use the 
relationship (\ref{eq14}), given at the end of next section.

\section{The Probability Distribution, $P(\delta,r)$. Most Probable and Mean Profiles}\label{sec2}

At a given value of $r$, $\delta$ takes different values, $\delta(r,j)$, over the assemble 
of halos. The question now is which is the probability distribution of $\delta$ over this ensemble, P($\delta$,r). 
As we saw in the previous section, this can be done, in principle, by making realizations of the initial 
profile, $\delta_{l}(q,i)$, and evolving them accordingly with equations (\ref{eq3}). Let's assume, as a 
first approximation, that the realizations of the initial profile may be carried out by generating 
for each value of q a value of $\delta_{l}$ accordingly with distribution (\ref{eq4}). That is, we assume 
that the distribution of $\delta_{l}$ is only conditioned by the fact that $\delta_{l}(Q)=\delta_{vir}$. 
We shall latter consider initial profiles with an additional constraint. With these realizations we 
can elaborate for each value of r a histogram for $P(\delta,r)$. There is, however, a direct analytical 
procedure to obtain $P(\delta,r)$ from the probability distribution for $\delta_{l}$ (expression (\ref{eq4}) 
in the present approximation). 

$\delta$ is a unique function of $\delta_{l}$ (expression (\ref{eq2})). So, one may think that P($\delta$,r) can be 
obtained from (\ref{eq4}) simply through the change of variable $\delta=\delta(\delta_{l})$. However, 
expressions (\ref{eq3}) show that in transforming the initial 
profile not only is $\delta_{l}$ transformed into $\delta$, but also $q$ is transformed into $r$. Now, 
expression (\ref{eq4}) gives the distribution of $\delta_{l}$ at a fixed $q$ value. But 
what we want to obtain is the distribution for $\delta$ at a fixed $r$ value. So, since the relationship 
between $q$ and $r$ depends on $\delta$ (or $\delta_{l}$) itself, it is clear that the derivation of 
P($\delta$,r) from expression (\ref{eq4}) (i.e. from P($\delta_{l}$,q)) can not be as simple as described above.

Fortunately, there is a simple expression relating both probability distributions, which is valid as 
long as shell-crossing is not important:

\begin{equation}
P(\delta,r)=-\frac{d}{d\delta}\frac{\int_{\delta_{l}(\delta)}^{\delta_{vir}} P(\delta_{l},q)~d\delta_{l}} 
{\int_{-\infty}^{\delta_{vir}} P(\delta_{l},q)~d\delta_{l}} \label{eq9} 
\end{equation}
\begin{displaymath}
q\equiv r~(1+\delta)^\frac{1}{3}    
\end{displaymath} 

where $\delta_{l}(\delta)$ is given by expression (\ref{eq1}) and $P(\delta_{l},q)$ is the linear profile. 
Note that $\delta$ enters not only in the integration limit but also in the integrand through $q$. The 
derivation of this relationship is given in Patiri et al.(2004) appendix B. In that work 
this relationship was derived in regard with void density profiles, so it had an slightly different form. 
In appendix B we give the derivation corresponding to the present case. 

Using expression (\ref{eq4}) for P($\delta_{l}$,q) we find:

\begin{equation}
\frac{\int_{\delta_{l}(\delta)}^{\delta_{vir}} P(\delta_{l},q)~d\delta_{l}} 
{\int_{-\infty}^{\delta_{vir}} P(\delta_{l},q)~d\delta_{l}} = 
\frac{1}{2}~\frac{erfc(F(x=\delta_{l}(\delta)))-erfc(F(x=\delta_{vir}))}{1-\frac{1}{2}erfc(F(x=\delta_{vir}))} \label{eq9b}
\end{equation}

\begin{displaymath}
F(x) \equiv \frac{x-\frac{\sigma_{12}(q)}{\sigma(Q)}~\delta_{vir}}{\sqrt{2~g(q)}}
\end{displaymath}

\begin{displaymath}
q\equiv (r~(1+\delta))^{\frac{1}{3}}  
\end{displaymath}

with $\sigma_{12}(q)$, $\sigma(Q)$, $g(q)$ as defined in (\ref{eq4}). 

For the purposes of this section we may neglect the term $erfc(F(x=\delta_{vir}))$. The full expression 
shall be used in section 5 along with a more refined one. We then have:

\begin{equation}
P(\delta,r)=\frac{1}{\sqrt{\pi}}~e^{-F^{2}(x=\delta_{l}(\delta))}\frac{d}{d\delta}F(x=\delta_l(\delta))     \label{eq10}
\end{equation}

By construction, $\delta$ at $r$ must be smaller than $\Delta_{vir}$ and larger than certain value, $\delta_{min}(r)$:

\begin{equation}
  \delta_{min}(r)=341~(R_{vir}/r)^3-1    \label{eq9c}
\end{equation}

This minimum value corresponds to a situation where there is no matter in between $R_{vir}$ and $r$. So, for $\delta$ 
values outside the interval $(\delta_{min}(r),\Delta_{vir})$, $P(\delta,r)$ is zero.

We may obtain an analytical expression for P($\delta$,r) using approximation (\ref{eq5}) 
for $\frac{\sigma_{12}(q)}{\sigma_{1}^{2}}$ and the following approximation for $\sigma(q)$ (which enters $g(q)$, 
defined below expression (\ref{eq4})):

\begin{displaymath}
  \sigma(q) \simeq (1.65~10^{-2}+0.105(q~h~Mpc^{-1}))^{-\frac{1}{2}}   \qquad  \textrm{for }q < 3~h^{-1}Mpc
\end{displaymath}

As mentioned before, expression (\ref{eq9}) is valid as long as shell-crossing is not important. 
In Prada et al.(2005) we have found by mean of comparison with numerical simulations that, beyond three 
virial radius, the relevance of shell-crossing diminishes quickly. This relevance can be estimated 
a priori (i.e. without comparison with simulations) obtaining P($\delta$,r) directly through realizations 
of the initial profile accordingly with expression 
(\ref{eq4}) and evolving them accordingly with equations (\ref{eq3}). If shell-crossing were irrelevant, 
the P($\delta$,r) obtained in this way should be equal to that given by expression (\ref{eq10}). The 
presence of certain amount of shell-crossing will cause the P($\delta$,r) obtained with realization to 
have a somewhat smaller maxima and a more extended tail than that given by (\ref{eq10}) (note that it is this 
expression that corresponds to realizations elaborated with expression (\ref{eq4}). In this case 
none of the procedures gives the correct P($\delta$,r) because both assume that $\delta$ is related to 
$\delta_{l}$ by means of expression (\ref{eq1}), which is inconsistent when shell-crossing is important. 
However, the difference between the results obtained with both procedures is of the same order of the 
difference between any of them and the profile obtained with a proper treatment of shell-crossing. Note 
that even this last P($\delta$,r) is not the real one, since, as we said before, we are generating the 
initial profile using only a two point distribution (expression (\ref{eq4})).

\begin{figure}[!h]\centering
\includegraphics[width=8.75cm,height=7.5cm]{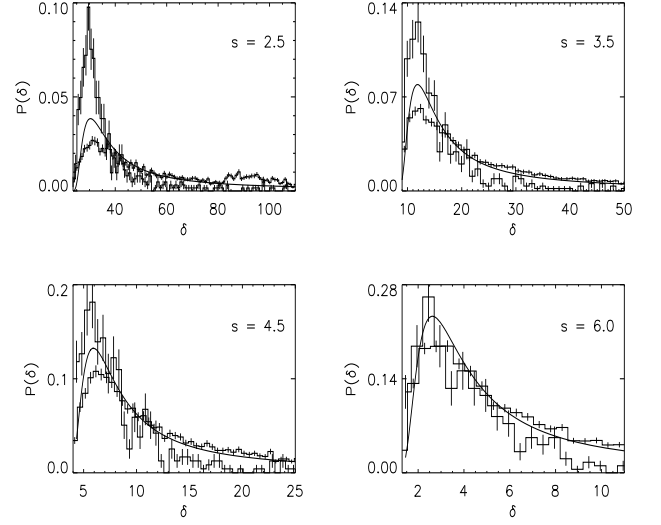}
\caption{$P(\delta,r)$ as given by expression (\ref{eq10}) compared with the corresponding histogram obtained
through realizations (histogram with small error bars) and that found in the simulations (histogram with 
large error bars) for four values of $s~(\equiv r/R_{vir})$ and a mass of 
$3\times10^{12}\msunh$ ($\Delta_{vir}=340$ and $\delta_{vir}=1.9$).}
\label{fig1}
\end{figure}

In figure \ref{fig1} we compare the P($\delta$,r) obtained by the two procedures mentioned above
for several values of $r$ expressed in unit of the virial radius (that we denote by s). We also show 
the corresponding histograms obtained from the numerical simulations described in Prada et al.(2005), which 
were done using the Adaptive Refinement Tree (ART) code (Kravtsov et al. 1997) for the standard 
$\Lambda$CDM cosmological model with $\Omega_{0}=0.3$, $\Omega_{\Lambda}=0.7$, $h=0.7$, and
cover a wide range of scales with different mass and force resolutions (see Prada et al. 2005 for a 
detailed description). In particular, the histograms in figure 1 were obtained for a total of 654 halos in a mass 
range of $3 \pm 1 \cdot 10^{12} h^{-1}~M_{\sun}$ selected without any kind of isolation criteria.

We can see that, for $s=2.5$, where shell-crossing is already important, there is substantial difference 
between the results of both procedures. A considerable amount of probability is transfered from the 
most probable value to much larger values ($\delta \approx 100$) causing the distribution obtained 
through the realizations to be bimodal. For $s=3.5$ there is still a small amount of shell-crossing 
causing the maxima obtained with both procedures to differ by roughly a 20\%. For larger values of $s$ 
this difference steadily diminishes.

Furthermore, as $s$ increases the difference between the distribution given by expression 
(\ref{eq10}) and the histogram obtained with the numerical simulations reduces. In fact, even 
for $s=3.5$ the relative values of P($\delta$,r) at different values of $\delta$ to the left 
of the maxima are very well given by (\ref{eq10}). The difference in the absolute values with 
respect to those in the histogram is due to the normalization. For $\delta$ values to the right 
of the maxima, expression (\ref{eq10}) gives a considerably more extended tail than in the actual 
distribution. Therefore the normalization constant is larger in the latter case.

Note that, although the maxima of expression (\ref{eq10}) and that of the histogram corresponding 
to the numerical simulations approach as $s$ increases, due to the increasing irrelevance of the tail, 
this tail is still substantially more extended even for $s=6$. 
Now, since the relevance of shell-crossing is small for $s$ larger than 3, the most 
likely explanation for the excess in the tail given by expression (\ref{eq10}) lies on the fact that 
we are using expression (\ref{eq4}) for P($\delta_{l}$,q). In the last section we shall consider a 
better P($\delta$,r) and discuss the resulting improvement of the behaviour of its tail. 

Using P($\delta$,r) (as given by (\ref{eq10})) we may immediately obtain the most probable and 
the mean profiles. For the first one we have:

\begin{equation}
\delta_{p}(r) \equiv \delta_{max} ; \qquad \frac{d}{d\delta}P(\delta,r)\bigg|_{\delta=\delta_{max}}=0   \label{eq11}
\end{equation}

And for the mean profile, in principle:

\begin{equation}
\overline\delta(r) \equiv \int_{\delta_{min}}^{\Delta_{vir}}P(\delta,r)\delta~d\delta  \label{eq12}
\end{equation} 

However, due to the fact that the mean is rather sensitive to the form of the tail, we must artificially 
cut off the tail. From the simulations 
we know that the real tail practically ends at $\delta \thicksim \delta_{0}(r)$ with $\delta_{0}$ given by:

\begin{displaymath}
P(\delta_{0},r)=\frac{P(\delta_{max},r)}{25}
\end{displaymath}

So, instead of (\ref{eq12}) we use (Prada et al.2005):

\begin{equation}
\overline\delta(r) \equiv 
\frac{\int_{\delta_{min}}^{\delta_{0}(r)}P(\delta,r)\delta~d\delta}{\int_{\delta_{min}}^{\delta_{0}(r)}P(\delta,r)~d\delta}    
\label{eq13}
\end{equation} 

In table 1 we give the values of $\delta_{0}$ and $\delta_{max}$ for several values of $s~(\equiv r/R_{vir})$.

So far we have considered the probability distribution and profiles of the spherically-averaged enclosed density 
contrast, $\delta$. We shall now consider the local density contrast $\delta$', that is the density 
contrast within a narrow shell of radius $r$. In this case we can not obtain P($\delta$',r) by mean 
of a simple relationship like expression (\ref{eq9}). However, the mean $\delta$' profile can
be obtained from the mean $\delta$ profile. To this end, note that for each actual density profile, that is, 
for each spherically evolved realization of the linear profile, $\delta(r,j)$, the following 
relationship holds:

\begin{equation}
\delta'(r,j) = \frac{1}{3r^{2}}~\frac{d}{dr}(r^{3}\delta(r,j))    \label{eq14}
\end{equation} 

which follows immediately from the definitions of $\delta$, $\delta'$. The mean $\delta$ profile 
does not correspond to any actual density profile. However, being the mean a linear operation, the same 
relationship holds for the mean profiles:

\begin{eqnarray}
\overline\delta'(r) \equiv <\delta'(r,j)>_{j} = <\frac{1}{3r^{2}}~\frac{d}{dr}(r^{3}\delta(r,j))>_{j} = \\ \nonumber
=\frac{1}{3r^{2}}~\frac{d}{dr}(r^{3}<\delta(r,j)>_{j}) = \frac{1}{3r^{2}}~\frac{d}{dr}(r^{3}\overline\delta(r))   
\label{eq15}
\end{eqnarray}

\begin{table}[!h] \centering
\caption{Artificial cut-off for $P(\delta,r)$}
\label{table1} 
\vspace{0.2 cm}
\begin{tabular}{ccc}  

\hline  \hline
s & $\delta_{max}$ & $\delta_{0}$   \\
\hline  
\noalign{\smallskip} 
1.5 & 115.2 & 495.4   \\
2.5 & 28.6 & 135.3   \\
3.5 & 11.3 & 67.9    \\
4.5 & 5.8 & 42.2    \\
5.5 & 3.3 & 29.3    \\
6.5 & 2.0 & 19.6    \\
7.5 & 1.3 & 12.8    \\
8.5 & 0.82 & 8.1     \\
\hline

\end{tabular}
\end{table}

The most probable profile is constructed with the ensemble of profiles $\delta(r,j)$ by means 
of a non-linear operation: that of choosing, for any value of r, the member of the ensemble 
with the largest probability density for $\delta(r)$. So, relationship (\ref{eq14}) is not valid between 
$\delta'_{p}$ and $\delta_{p}$, because the value of $\delta_{p}$ at different 
values of r may correspond to different members of the ensemble. However, in Prada et al.(2005) 
we have used expression (\ref{eq14}) to obtain $\delta'_{p}$(r) from $\delta_{p}(r)$ and 
found results in good agreement with the simulations, but this is a posteriori agreement: 
unlike the prediction for $\overline\delta'(r)$, the value of $\delta'_{p}$(r) obtained 
using expression (\ref{eq14}) is not a proper prediction.

\section{Radial velocity profile of dark matter halos} \label{Sec3}

In the spherical collapse model, the radial velocity at distance $r$ with respect to the center of 
the spherical cloud is a unique function of the spherically-averaged enclosed density contrast, $\delta(r)$. 
An exact analytical expression may be given for this relationship using $\delta_{l}(\delta)$ 
(expression (\ref{eq1})). Mass conservation within a shell with initial Lagrangian 
radius $q$, implies that, at any time, the following relationship must hold:

\begin{equation}
\frac{r(t)}{a(t)} = q(1+\delta)^{-\frac{1}{3}}    \label{eq16}
\end{equation} 

where $a(t)$ is the scale factor of the universe (normalized to 1 at present) so that the left 
hand side is the comoving radii of that shell at time $t$. Deriving this equation with respect 
to time we find:

\begin{equation}
\frac{\dot{r}(t)}{a(t)}-\frac{\dot{a}(t)}{a^{2}(t)}~r(t) = \frac{\dot{r}(t)-H(t)~r(t)}{a(t)} = 
-\frac{1}{3}~q(1+\delta)^{-\frac{4}{3}}~\dot{\delta}  \label{eq17}
\end{equation} 

where $H(t)$ is the Hubble constant. Using now $\delta(\delta_{l})$ we may express $\dot{\delta}$ in the form:

\begin{equation}
\dot{\delta} = \frac{d\delta(\delta_{l})}{d\delta_{l}}~\dot{\delta}_{l} ; \qquad \dot{\delta}_{l} = 
\frac{\dot{D}(t)}{D(t)}~\delta_{l} = \frac{a(t)}{D(a(t))}\frac{dD(a(t))}{da(t)}~H~\delta_{l}    \label{eq18}
\end{equation} 

where $D(t)$ is the growing mode of linear density fluctuations. Now, since $\delta(\delta_{l})$ is the 
inverse function of $\delta_{l}(\delta)$ we have:

\begin{displaymath}
\frac{d\delta(\delta_{l})}{d\delta_{l}} = \left(\frac{d\delta_{l}(\delta)}{d\delta}\right)^{-1}
\end{displaymath}

Using this in (\ref{eq18}), inserting (\ref{eq18}) in (\ref{eq17}) and using again (\ref{eq16}) we find:

\begin{equation}
V_{r} \equiv \dot{r} = 
H(t)~r-\frac{1}{3}\frac{a(t)H}{D(a(t))}~\frac{dD(a(t))}{da(t)}~\frac{r}{(1+\delta)}~\frac{\delta_{l}(\delta)}  
{\frac{d\delta_{l}(\delta)}{d\delta}} \equiv r~f(\delta)   \label{eq19}
\end{equation}

For the concordant cosmology we have at present $H=72~ km/s/Mpc$, $a=1$, $\frac{dD(a)}{da}/D(a)=0.51$. 
Writing (\ref{eq19}) in the form $V_{r}=rf(\delta)$ we may use expression (\ref{eq10}) to obtain the 
probability distribution, $P(V_{r},r)$, for $V_{r}$ at radii $r$:

\begin{equation}
P(V_{r},r) = P\left(\delta=f^{-1}\left(\frac{V_{r}}{r}\right),r\right)~\left(\frac{df(\delta)}{d\delta}\right)^{-1}  
~\frac{1}{r}
\label{eq20}
\end{equation}

where $f^{-1}$ is the inverse function of $f$ and where $P(\delta,r)$ is given by expression (\ref{eq10}).

This distribution should not be confused with the distribution for the radial velocity of dark matter particles at a given value $r$. 
$V_{r}$ is the mean radial velocity  of all particles in a given narrow shell with radius $r$. So, for a 
given halo and a given value of $r$, $V_{r}$ takes a unique value. Expression (\ref{eq20}) gives the 
distribution of this value over the ensemble of halos.

The mean $V_{r}$ profile is given by:

\begin{equation}
\overline V_{r} = 
\frac{\int_{\delta_{min}}^{\delta_{0}(r)}P(\delta,r)rf(\delta)~d\delta}{\int_{\delta_{min}}^{\delta_{0}(r)}P(\delta,r)~d\delta}   
\label{eq21}
\end{equation}

with $P(\delta,r)$ given by (\ref{eq10}), $\delta_{0}$ as given in Table 1, $f(\delta)$ given by (\ref{eq19}) and $\delta_{min}$ 
given by (\ref{eq9c}).

\begin{figure}[!h] \centering
\includegraphics[width=8.75cm,height=7.5cm]{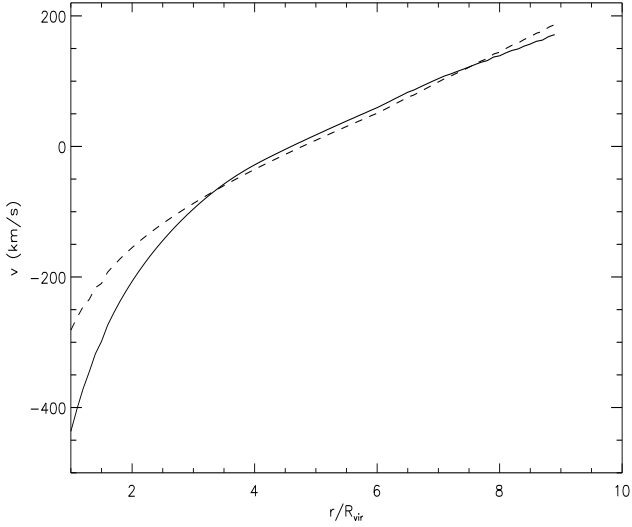}
\caption{Radial velocity profile as given by expression (\ref{eq21}) (filled line) and as given by expression  
(\ref{eq22}) (dashed line). Both correspond to a mass of $3 \times 10^{12}~h^{-1}M_{\sun}$}
\label{fig2}
\end{figure}

In figure \ref{fig2} we show the predictions given by (\ref{eq21}) for the mean radial velocity profile, and compare it with the profile:
\begin{equation}
V_{r} \equiv r~f(\overline\delta(r))   \label{eq22}
\end{equation}

That is, for each value of $r$, this expression gives $V_{r}$ corresponding to the mean $\delta$ at that $r$ through Eq.(\ref{eq19}). We find that Eq.(\ref{eq22}) is a good approximation to Eq.(\ref{eq21}).

Note that, although we have used the probability distribution given by Eq.(\ref{eq10}), these expressions are valid for any $P(\delta,r)$.

\section{Improving $P(\delta,r)$}

We have seen that for $s$ ($=r/R_{vir}$) larger than 3 the realization of initial profiles and their subsequent 
evolution lead to values of $P(\delta,r)$ very close to those obtained using expression (\ref{eq10}). This means that 
shell crossing is not important at these radii. So, the difference between the actual value of 
$P(\delta,r)$ (the histogram obtained from the simulations) and that given by expression (\ref{eq10}) lies on the fact 
that this expression is only an approximation, or in the possible relevance of triaxiality, non-radial motions and pressure 
(velocity dispersions). To determine the amount of the discrepancy due to inaccuracies in the statistical description 
of the initial conditions as opposed to the discrepancy due to neglected dynamical factors, we must consider better 
approximation than that provided by expression (\ref{eq10}). We shall use now two improved approximations. The first one 
is that obtained by using expression (\ref{eq9b}) in expression (\ref{eq9}), which we represent by $P_1(\delta,r)$. Note 
that, to derive this approximation we have used the fact that at a given $q$ larger than $Q$, $\delta_l$ must be smaller 
than $\delta_{vir}$, but we have used expression (\ref{eq4}) for $P(\delta_{l},q)$. This distribution does not account 
for the fact that for all values of q larger than Q, $~\delta_{l}<\delta_{vir}$. 
Consequently, $P(\delta_{l},q)$ falls off with $\delta_{l}$ more slowly than it should and the same 
applies to $P(\delta,r)$. In the second approximation, which we represent by $P_2(\delta,r)$, we consider a 
$P(\delta_{l},q)$ which takes into account this additional constraint. By comparing the predictions obtained with 
$P_2(\delta,r)$, $P_1(\delta,r)$ with those obtained with $P(\delta,r)$ (expression (\ref{eq10})) we may check 
whether $P_1(\delta,r)$ is accurate enough, so that the remaining discrepancy of the prediction with the results 
shown in the simulations may be ascribed to unaccounted dynamical factors.

We have already pointed out that the most relevant part of the present density profile (up to $\thicksim$ 8 virial  
radii) comes from a narrow band in Lagrangian coordinates (from $Q$ to $\thicksim$ $1.5Q$) so that the values 
of $\delta_{l}$ within this band are strongly correlated. This is the reason why using expression 
(\ref{eq4}) for $P(\delta_{l},q)$ is a good approximation, because if $\delta_{l}$ at $q$ (within the mentioned band) lies below 
$\delta_{vir}$ the same will probably be valid also at all $q'$ larger than $Q$. We go now a step further 
and explicitly demand that $\delta_{l}$ lies below $\delta_{vir}$ at all Lagrangian radii in between 
$Q$ and $q$. Imposing the same condition at $q'$ larger than $q$ is unnecessary since it will lead to a negligible change  
for $P(\delta_{l},q)$, because if $\delta_{l}$ lies below $\delta_{vir}$ at $q$ the same will almost certainly 
occur at larger distances.

Given the strong correlation between the values of $\delta_{l}$ in between $Q$ and $q$ (for the relevant $q$'s), 
the condition that for all values of q larger than Q $~\delta_{l}<\delta_{vir}$, is almost equivalent to demanding  
that $\delta_{l}$ lies below $\delta_{vir}$ 
at the middle point $q'=\frac{1}{2}(q+Q)$. We might have chosen any other point in between and 
searched for the point imposing the strongest constraint, since the real constraint must be stronger 
(i.e. the tail of the distribution falls off more steeply) than that imposed by the point leading to 
the strongest constraint. We have chosen the middle point because it seems a priori a good choice.

Here we discuss the main lines of the derivation and the result for $P(\delta_{l},q)$ within this new condition, which  
we represent by $P_{2}(\delta_{l},q)$. We give in appendix \ref{apendA} the details and the full descriptions of the 
expressions involved.

\begin{figure}[!ht]  \centering
\includegraphics[width=8.75cm,height=7.5cm]{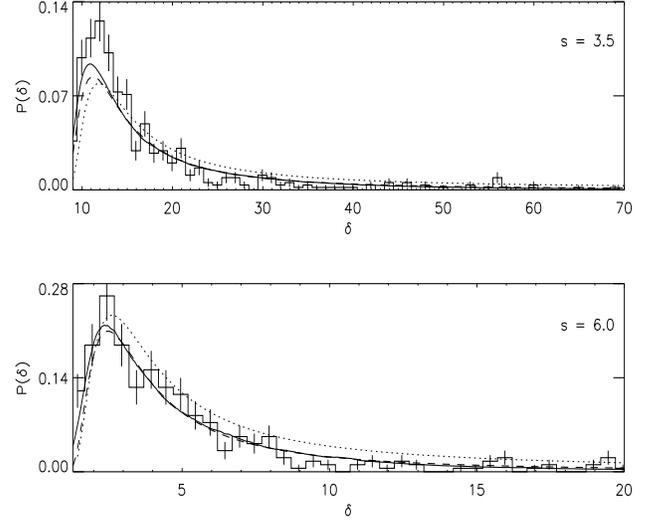}
\caption{Probability distribution for $\delta$ at 3.5 and 6 virial radius for $3 \times 10^{12}~h^{-1}M_{\sun}$. The  
filled curve corresponds to the approximation given by expression (\ref{eq24}), the dashed curve to that given by 
full expression (\ref{eq9}), and the dotted curve to expression (\ref{eq10}). The histogram corresponds to the same simulations 
as in fig.\ref{fig1}.}
\label{fig3}
\end{figure} 

\begin{figure}[!h]  \centering
\includegraphics[width=8.75cm,height=7.5cm]{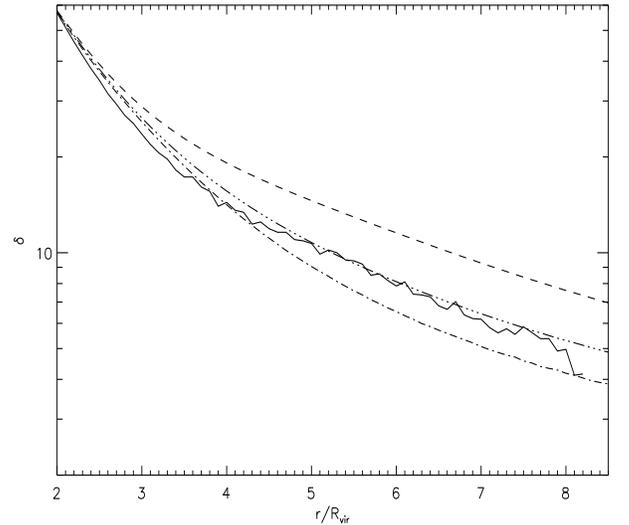}
\caption{Mean $\delta$ profile for $3 \times 10^{12}~h^{-1}M_{\sun}$ using the probability distribution given by 
expression (\ref{eq10}) (dashed line), by full expression (\ref{eq9}) (3dots-dashed line) and by expression (\ref{eq24}) 
(dot-dashed line). Mean $\delta$ obtained from simulations is given for comparison (filled line). In all the cases, 
a maximum value of $\delta=70$ was used.}
\label{fig4}
\end{figure} 

\begin{figure}[!h]  \centering
\includegraphics[width=8.75cm,height=7.5cm]{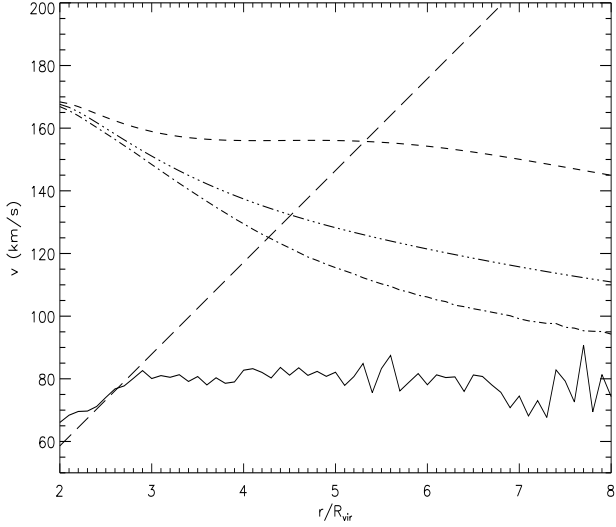}
\caption{Peculiar infall velocity profile for $3 \times 10^{12}~h^{-1}M_{\sun}$ using the probability distribution given by 
expression (\ref{eq10}) (dashed line), by full expression (\ref{eq9}) (3dots-dashed line) and by expression (\ref{eq24}) 
(dot-dashed line). Peculiar infall velocity profile obtained from simulations is given for comparison (filled line). 
In all the cases, a maximum value of $\delta=70$ was used. The straight line corresponds to the Hubble Flow.}
\label{fig5}
\end{figure} 

\begin{figure}[!h]  \centering
\includegraphics[width=8.75cm,height=7.5cm]{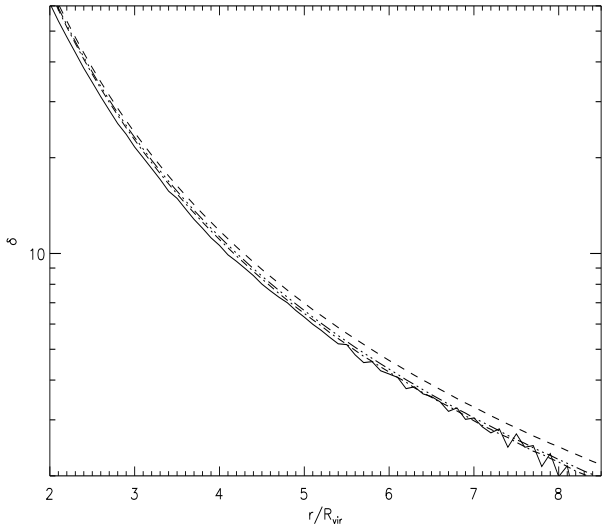}
\caption{Mean $\delta$ profile for $3 \times 10^{12}~h^{-1}M_{\sun}$ using the probability distribution given by 
expression (\ref{eq10}) (dashed line), by full expression (\ref{eq9}) (3dots-dashed line) and by expression (\ref{eq24}) 
(dot-dashed line). Mean $\delta$ obtained from simulations is given for comparison (filled line). For all radius, 
the average was calculated excluding the 20\% of the halos with the largest $\delta$ values.}
\label{fig6}
\end{figure} 

\begin{figure}[!h]  \centering
\includegraphics[width=8.75cm,height=7.5cm]{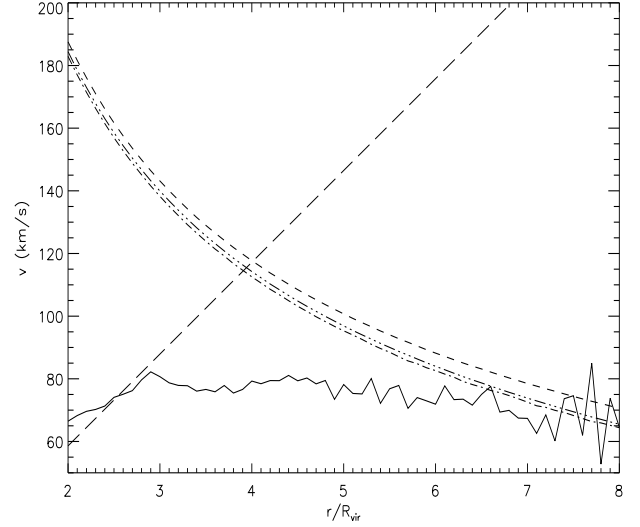}
\caption{Peculiar infall velocity profile for $3 \times 10^{12}~h^{-1}M_{\sun}$ using the probability distribution given by 
expression (\ref{eq10}) (dashed line), by full expression (\ref{eq9}) (3dots-dashed line) and by expression (\ref{eq24}) 
(dot-dashed line). Peculiar infall velocity profile obtained from simulations is given for comparison (filled line).
For all radius, the average was calculated excluding the 20\% of the halos with the largest $\delta$ values. The straight 
line corresponds to the Hubble Flow.}
\label{fig7}
\end{figure} 

To obtain $P_{2}(\delta_{l},q)$ we must first obtain the joint probability distribution for the value 
of $\delta_{l}$ at $Q$, the middle point, and at $q$. We represent by $x_{1}$, $x_{2}$, $x_{3}$ respectively the value  
of $\delta_{l}$ at these three points.

The joint distribution for these three variables is a Gaussian trivariate distribution $P(x_{1},x_{2},x_{3})$, 
which can be obtained for a given power spectra. With this distribution we may 
immediately obtain the distribution of $x_{3}$ conditioned to $x_{1}=\delta_{vir}$, $x_{2}<\delta_{vir}$, 
namely: $\overline{P_{2}}(\delta_{l},q)$. So, we have:

\begin{equation}
\overline{P_{2}}(\delta_{l},q) = 
\frac{\int_{-\infty}^{\delta_{vir}}P(x_{2}|x_{1})~P(x_{3}|x_{1},x_{2})~dx_{2}}{\int_{-\infty}^{\delta_{vir}}P(x_{2}|x_{1})~dx_{2}} 
\label{eq23}
\end{equation}

where $P(\delta_{vir},x_{2})$ is the joint probability distribution for $x_{1}$, $x_{2}$ 
with $x_{1}=\delta_{vir}$. $\overline{P_{2}}(\delta_{l},q)$ is the probability distribution of $\delta_l$ at $q$ 
conditioned to $\delta_l(Q)=\delta_{vir}$ and $\delta_l(q')<\delta_{vir}$. We have not imposed yet the almost redundant 
condition (since $\overline{P_{2}}(\delta_{l},q)$ is very small for $\delta_l>\delta_{vir}$) that $\delta_l$ at $q$ 
must be smaller than $\delta_{vir}$. The probability distribution for $\delta_l$ with this condition, 
$P_{2}(\delta_{l},q)$, is simply obtained through normalization (see (\ref{eqA5})) 

The distribution for $\delta$ at fixed $r$ within this formalism, which we represent by $P_{2}(\delta,r)$, 
may be obtained from $P_{2}(\delta_{l},r)$ by means of expression (\ref{eq9}), which is valid when 
shell-crossing is not important ($s \gtrsim 3$):

\begin{equation}
P_{2}(\delta,r)=-\frac{d}{d\delta}\int_{\delta_{l}(\delta)}^{\delta_{vir}}P_{2}(\delta_{l},q)~d\delta_{l}   \label{eq24}
\end{equation}
\begin{displaymath}
q \equiv r(1+\delta)^{\frac{1}{3}}
\end{displaymath}

In figure \ref{fig3} we show the probability distributions for a mass of $3 \times 10^{12}~h^{-1}M_{\sun}$ 
at 3.5 and 6.0 virial radius. As expected, $P_{2}(\delta,r)$ falls off much faster than $P(\delta,r)$ being
in excellent agreement with the simulations. The tail of $P_1(\delta,r)$ falls off sufficiently fast to give sensible results for the 
density and velocity profiles averaged over all possible halos ($\delta$ between $\delta_{min}$ and $\Delta_{vir}$). 
However, we know that for $\delta \gtrsim 70$ the standard spherical collapse model is not a good approximation, so we 
can not learn much by comparing simulations with predictions for averages over all halos. It is more instructive to 
compare the predictions for the averages corresponding to $\delta$ values between $\delta_{min}$ (expression (\ref{eq9c})) 
and $70$ with those found in the simulations for the same range of $\delta$ values.

In figure \ref{fig4} we show the predictions given by $P(\delta,r)$, $P_1(\delta,r)$, $P_2(\delta,r)$ for the $\delta$ 
profile averaged between $\delta_{min}$ and $70$ and the averaged found in the simulations. In figure \ref{fig5} the corresponding 
results for the peculiar infall velocity (the Hubble Flow minus the radial velocity) are presented. One of the things we 
learn from these figures is that $P_2(\delta,r)$ must differ very little from the exact distribution, since it may be shown on 
general grounds that the difference between the predictions given by these two distributions must be much smaller than the 
difference between the predictions given by $P_2(\delta,r)$ and $P_1(\delta,r)$. Another thing that we learn is that the spherical 
collapse model can not be a good approximation for all halos included in the averaging process. The reason being that only 
the presence of some considerably triaxial haloes (i.e. haloes such that the density contours of their outskirts are very triaxial) 
within the averaging assemble can explain the fact that the results found 
in the simulations are somewhat larger than those obtained using $P_2(\delta,r)$. If the dynamical factor that we have 
neglected were irrelevant the latter results must be slightly above the former. Non-radial motion and velocity dispersions 
preserve the order, only increasing the difference. The reason being that when these factors are taken into account, while 
preserving spherical symmetry, $\delta_l(\delta)$ is steeper ($\delta$ smaller for given $\delta_l$) than in the standard model 
(expression (\ref{eq1})), resulting, through expression (\ref{eq24}) and (\ref{eqA8}) in steeper $\overline{\delta}(r)$. 
So, only the relevance of triaxiality can explain the results shown in figure \ref{fig4}. The effect is not large in 
this figure, since only a small fraction of the halos are affected, but it is 
quite meaningful. It is this very effect that is causing that the peculiar infall velocity profile predicted by means of 
$P_2(\delta,r)$ does not agree with that found in the simulations even at $r/R_{vir} \gtrsim 7$.

Knowing that for some halos the spherical collapse model can not be a good approximation, the relevant question now is to 
determine precisely how good is it for most halos. To this end, at each radial bin we search for the value, $\delta_1(r)$, 
such that the upper cumulative probability, as given by $P_2(\delta,r)$, is $0.2$, and eliminate, both in the simulations 
and in the predictions, halos with larger $\delta$ values at that bin. By doing so, we eliminate many halos which simply 
have flatter profiles than average, but are otherwise sufficiently spherical for the model to apply. 
But, we are sure of having eliminated all halos with highly triaxial outskirts, most of them corresponding to situations 
where a couple of halos lie within a few virial radius 
from each other, so that each of them will show up in the outskirt of the other halo; a situation that, by no mean, can be 
described by the spherical model. The resulting profiles, as given by the different approximations to $P(\delta,r)$ and the 
simulations are given in figure \ref{fig6}. The difference between the various approximations is now smaller, since the main 
difference occurs at the far tail (of $P(\delta,r)$), which has now been eliminated, 
but this difference increases for smaller masses. The 
profile obtained from the simulations lies now slightly below the best prediction, as it should be if the spherical model 
is a good approximation. In figure \ref{fig7} we show the corresponding results for the peculiar infall velocity. It is 
apparent that the prediction agrees well with the simulations at large radii but as we go below $\thicksim$ 5 virial radius there 
is an increasing discrepancy. This must be due to the unaccounted dynamical factors. 

\begin{figure}[!h]  \centering
\includegraphics[width=8.75cm,height=7.5cm]{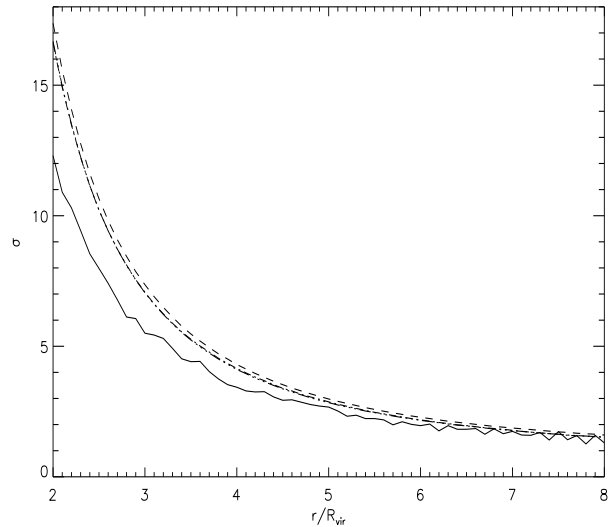}
\caption{$\sigma$ corresponding to the mean $\delta$ profiles (excluding at each radius the upper 20\% $\delta$ values) 
obtained using different probability distributions, as shown in fig.\ref{fig6} (same type of lines used here); 
$\sigma$ of the mean $\delta$ obtained from simulations is also given for comparison (filled line).}
\label{fig8}
\end{figure}

Figure \ref{fig8} give us another clue as to the relative effect of these factors. Triaxiality causes the dispersion of $\delta$ 
at any given radii to increase, since, for a given value of $\delta_l$ triaxial evolution gives a distribution of $\delta$ values 
with a finite dispersion, while the spherical model gives a single value. So, the fact that the dispersion found in the 
simulations lies somewhat below the predictions, imply that triaxiality can not be dominant amongst unaccounted factors.

In summary, the distribution $P_2(\delta,r)$ is very close to the exact. To most purposes the simpler distribution $P_1(\delta,r)$ 
may be used, their difference being small, although it increases for smaller values. The density profiles obtained with 
$P_2(\delta,r)$ are in excellent agreement with those found in simulations beyond $\thicksim$ 3 virial radius. The density 
dispersion and the radial velocity show some discrepancy below $\thicksim$ 5 virial radius, which clearly indicate the 
relevance of unaccounted dynamical factors, specially velocity dispersions.

\section{Final remarks}

The spherical collapse model describes very well the properties of dark matter halos beyond three virial radius. 
This could seem surprising given the fact that the density contours around halos may be considerably aspherical
and the presence of tidal fields. 

Nevertheless, the assumption that the spherically averaged density profiles at several virial radius evolves according
with the spherical collapse model is not just a simplification introduced to make the problem tractable. The mean  
evolution is given by the spherical collapse model with some dispersion (for a given $\delta_l$)
due to triaxiality. This have been shown by mean of simulations (Lilje et al. 1986) and analytical works 
(Bernardeau 1994). At sufficiently large radii where $\delta$ is sufficiently small ($\delta \lesssim 10$)
the fractional dispersion becomes small. At small radii, not only the dispersion becomes larger, causing the mean value 
of $\delta$ for a given $\delta_l$ to be somewhat different from $\delta(\delta_l)$ (expression (\ref{eq1})), 
but also the effect of non-radial motions and velocity dispersions starts to dominate. So studying the dark matter 
profile where the uncertainty of the evolution is small, we may check that the initial conditions 
we use are correct to a high degree of accuracy. With these conditions, as described by the most accurate
probability distribution, $P_{2}(\delta_{l},q)$, we should be able to obtain very accurate predictions for all  
possible definitions (typical, mean...) of density and velocity profile. That is, there is no room for remodeling; all properties 
must be explained with one and the same $P_{2}(\delta_{l},q)$. Any residual discrepancy should be
explained by triaxiality, non-radial motions and velocity dispersions, as we shall show in a future work, although, from 
the results shown in this work, the last two effects dominate at least up to roughly 5 virial radius.
Going a step further, if we adequately take into account the neglected dynamical factors, the same initial 
conditions should explain the profile at any radii. In this way, we could  
be able to explain the dark matter profiles at least down to a virial radii understanding the role played by the  
initial conditions and by the different processes relevant to the evolution.

\section*{acknowledgments}

M.A. thanks support from the Spanish MEC under grant PNAYA 2002-01241. We also thank support from MEC under grant PNAYA 2005-07789.

\appendix
\section{Derivation of $P_{2}(\delta_{l},q)$}  \label{apendA}

The field of initial density fluctuations linearly extrapolated to the present $\delta_{l}(\vec{q})$ is a random  
Gaussian field. Any set of quantities obtained through a linear operation on a Gaussian field follows a Gaussian  
multivariate distribution. The quantities we are interested in are the average of the linear field within three  
concentric spheres centered at a randomly chosen point with Lagrangian coordinate $\vec{q}$. The Lagrangian radius of  
these spheres are Q, $\frac{1}{2}(Q+q)$, q (note that q is not the norm of $\vec{q}$) and we represent the average of  
the linear field within them by $x_{1}(\vec{q})$, $x_{2}(\vec{q})$, $x_{3}(\vec{q})$ respectively. These are three  
random Gaussian fields, but their values at a randomly chosen $\vec{q}$ we represent simply by $x_{1}$, $x_{2}$,  
$x_{3}$. Their joint distribution is a Gaussian tri-variate:

\begin{displaymath}
P(x_{1},x_{2},x_{3}) = \frac{e^{-\frac{1}{2}\chi}}{(2\pi)^{\frac{3}{2}}\big| det~\bf{C}\big|}
\end{displaymath}
\begin{displaymath}
\chi = \sum_{i,j=1}^{3}\left(\bf{C}^{-1}\right)_{jj}x_{i}~x_{j} ; \qquad (\bf{C})_{ij} \equiv <x_{i}~x_{j}>
\end{displaymath}

where $\bf{C}^{-1}$ is the inverse matrix of $\bf{C}$, whose diagonal elements are the variances of the x's and the  
other elements are their correlations.

Using the definitions of the probability distribution for $x_{2}$ conditioned to a given value of $x_{1}$,  
$P(x_{2}|x_{1})$, and of the distribution for $x_{3}$ conditioned to given values of $x_{1}$, $x_{2}$,  
$P(x_{3}|x_{1},x_{2})$, we may write:

\begin{equation}
P(x_{1},x_{2},x_{3}) = P(x_{1})~P(x_{2}|x_{1})~P(x_{3}|x_{1},x_{2})   \label{eqA1}
\end{equation}

Now, by grouping terms in $\chi$ in an appropriate manner we infer that:

\begin{displaymath}
P(x_{1}) = \frac{e^{-\frac{x_{1}^{2}}{2\sigma_{1}^{2}}}}{\sqrt{2\pi}~\sigma_{1}}
\end{displaymath}

\begin{displaymath}
P(x_{2}|x_{1}) = \frac{e^{-\frac{(x_{2}-Px_{1})^{2}}{2g}}}{\sqrt{2\pi}~g^{\frac{1}{2}}}
\end{displaymath}
\begin{displaymath}
g \equiv \sigma_{2}^{2}~(1-\overline{c}_{12}^{2}) ; \qquad P \equiv \overline{c}_{12}~\frac{\sigma_{2}}{\sigma_{1}}
\end{displaymath}

\begin{equation}
P(x_{3}|x_{1},x_{2}) = \frac{e^{-\frac{(x_{3}-Ax_{1}-Bx_{2})^{2}}{2~\sigma'^{2}}}}{\sqrt{2\pi}~\sigma'}  \label{eqA2}
\end{equation}

\begin{displaymath}
\sigma'^{2} \equiv 
\sigma_{3}^{2}\frac{(1+2\overline{c}_{12}~\overline{c}_{13}~\overline{c}_{23}-(\overline{c}_{12}^{2}+\overline{c}_{13} 
^{2}+\overline{c}_{23}^{2}))} {(1-\overline{c}_{12}^{2})}  ; \qquad A \equiv 
-\frac{\overline{c}_{12}~\overline{c}_{23}-\overline{c}_{13}}{1-\overline{c}_{12}^{2}}~\frac{\sigma_{3}}{\sigma_{1}} 
 ; \qquad B \equiv -\frac{\overline{c}_{12}~\overline{c}_{13}-\overline{c}_{23}}{1-\overline{c}_{12}^{2}}~\frac{\sigma_{3}}{\sigma_{2}}
\end{displaymath}

\begin{displaymath}
\overline{c}_{ij} \equiv \frac{<x_{i}~x_{j}>}{\sigma_{i}~\sigma_{j}} ; \qquad \sigma_{i} \equiv \sigma(q_{i})
\end{displaymath}

\begin{displaymath}
q_{1} = Q ; \qquad q_{2} = \frac{1}{2}~(Q+q) ; \qquad q_{3}=q
\end{displaymath}

$\sigma(q)$ is the linear RMS density fluctuation as a function of Lagrangian radii (see expression (\ref{eq4})).
$<x_{i}~x_{j}>$ is what we called $\sigma_{ij}$ in expression (\ref{eq4}), with $q_i$, $q_j$ in the place of $q$, $Q$:

\begin{equation}
\sigma_{ij}=\sigma_{ij}(q_i)=\frac{1}{2\pi^{2}}\int_{0}^{\infty}\mid\delta_{k}\mid^{2}W_{T}(q_ik)~W_{T}(q_jk)~k^2~dk   \label{eqA3}
\end{equation}

We may now obtain the distribution for $x_{3}$ conditioned to $x_{1}=\delta_{vir}$, $x_{2}<\delta_{vir}$:

\begin{equation}
P(x_{3}|x_{1}=\delta_{vir},x_{2}<\delta_{vir})=\frac{\int_{-\infty}^{\delta_{vir}}P(x_{2}|x_{1})~P(x_{3}|x_{1},x_{2})~ 
dx_{2}}{\int_{-\infty}^{\delta_{vir}}P(x_{2}|x_{1})~dx_{2}}  \label{eqA4}
\end{equation}

Note that $P(x_{1})$ cancels.

Using (\ref{eqA2}) and rearranging terms in the exponent we may write:

\begin{eqnarray}
P(x_{2}|x_{1})~P(x_{3}|x_{1},x_{2}) = \frac{1}{2\pi~g^{\frac{1}{2}}~\sigma'}~exp 
\left[-\frac{(x_{3}-Ax_{1})^{2}}{2~\sigma'^{2}}-\frac{P^{2}x_{1}^{2}}{2g}+\frac{(\frac{Px_{1}}{2g}+\frac{Bx_{3}}{2\sigma'^{2}}-\frac{ABx_{1}}{2\sigma'^{2}})^{2}}{\left(\frac{B^{2}}{2\sigma'^{2}}+\frac{1}{2g}\right)}\right] \times \nonumber\\
\times~ 
exp\left[-\left(\frac{B^{2}}{2\sigma'^{2}}+\frac{1}{2g}\right)~\left(x_{2}-\frac{(\frac{Px_{1}}{2g}+\frac{Bx_{3}}{2\sigma'^{2}}-\frac{ABx_{1}}{2\sigma'^{2}})}{\left(\frac{B^{2}}{2\sigma'^{2}}+\frac{1}{2g}\right)}\right)^{2}\right]  \nonumber
\end{eqnarray}

$x_{2}$ appears only in the second factor and both integrals in (\ref{eqA4}) can directly be expressed in terms of the error complementary function:

\begin{displaymath}
P(x_{3}|x_{1}=\delta_{vir},x_{2}<\delta_{vir}) = 
\frac{1}{2\sqrt{\pi}g^{\frac{1}{2}}~L^{\frac{1}{2}}~\sigma'}~exp\left[-\frac{(x_{3}-A\delta_{vir})^{2}}{2~\sigma'^{2}}-\frac{P^{2}\delta_{vir}^{2}}{2g}+\frac{U^{2}}{L}\right] \times 
\end{displaymath}
\begin{displaymath}
\times~\left[1-\frac{1}{2}~erfc\left((\delta_{vir}-\frac{U}{L})~L^{\frac{1}{2}}\right)\right] \times 
~\left[1-\frac{1}{2}~erfc\left(\frac{\delta_{vir}~(1-P)}{\sqrt{2}~g^{\frac{1}{2}}}\right)\right]^{-1}  
\end{displaymath}

\begin{displaymath}
U \equiv \frac{\delta_{vir}~P}{2g}+\frac{B~x_{3}}{2~\sigma'^{2}}-\frac{AB\delta_{vir}}{2~\sigma'^{2}}
\end{displaymath}
\begin{displaymath}
L \equiv \frac{B^{2}}{2~\sigma'^{2}}+\frac{1}{2g}
\end{displaymath}

$\overline{P_{2}}(\delta_{l},q)$, as defined in the text, is given by:
\begin{displaymath}
\overline{P_{2}}(\delta_{l},q)= P(x_3=\delta_l | x_1=\delta_{vir}, x_2<\delta_{vir})
\end{displaymath}

So, $P_{2}(\delta_{l},q)$, which is simply $\overline{P_{2}}(\delta_{l},q)$ with the restriction: $\delta_l<\delta_{vir}$ and 
the corresponding normalization, is given by:

\begin{equation}
P_{2}(\delta_{l},q) = \left \{ \begin{array}{ll}
\frac{P(x_{3}=\delta_{l}~|~x_{1}=\delta_{vir},x_{2}<\delta_{vir})}{\int_{-\infty}^{\delta_{vir}}   
P(x_{3}=\delta_{l}~|~x_{1}=\delta_{vir},x_{2}<\delta_{vir})} & \rm{if~\delta_l<\delta_{vir}}  \\
0 & \rm{if~\delta_l \ge \delta_{vir}}                \label{eqA5}
\end{array} \right.
\end{equation}

Note that the right hand side in the case $\delta_l>\delta_{vir}$ depends on $q$ through $\sigma_{i}$, $\overline{c}_{ij}$. 
These quantities may be computed directly with arbitrary precision evaluating the integral entering their definition 
(expression (\ref{eqA2})).  
However, to be able to obtain efficiently $P_{2}(\delta,r)$, we need accurate fits to these quantities as a function  
of q. It is very difficult, however, to obtain consistent approximation for these quantities: very small errors in  
$\overline{c}_{ij}$ may lead to inconsistent result, for example, negative values for $\sigma'^{2}$. So, it is more  
expedient to fit directly the quantities ($A$,$B$,$P$,$\sigma'^{2}$) where the $\sigma_{i}$, $\overline{c}_{ij}$  
enter. We find the following fits:

\begin{equation}
A = \left(-e^{-1.386~b~\left((\frac{q}{Q})^{2}-1\right)}\right)~\frac{\sigma_{3}}{\sigma_{1}} 
; \qquad B = \left(1+e^{-1.504~b~\left((\frac{q}{Q})^{2}-1\right)}\right)~\frac{\sigma_{3}}{\sigma_{2}} 
; \qquad P = e^{-0.475~b~\left((\frac{q}{Q})^{2}-1\right)}    \label{eqA6}
\end{equation}

\begin{displaymath}
\sigma'^{2} = 
\sigma_{3}^{2}~\frac{6.63~10^{-2}~b^{4}\left((\frac{q}{Q})^{2}-1\right)^{4}}{1-\left(\frac{\sigma_{1}}{\sigma_{2}}~P\right)^{2}}
\end{displaymath}

where b is as defined in expression (\ref{eq4}) ($b=0.2544$ for $M=3\times 10^{12}~h^{-1}M_{\sun}$). This fit is valid in the 
range $3\times 10^{11}$ to $3\times 10^{13}~h^{-1}M_{\sun}$. Outside this range it may be necessary using the full expressions 
(\ref{eqA2}) or a different fit to them.

Using expression (\ref{eqA5}) in expression (\ref{eq24}) we find:

\begin{equation}
  P_{2}(\delta,r) = -\frac{d}{d\delta}~G(\delta,q=r(1+\delta)^{\frac{1}{3}})       \label{eqA7}
\end{equation}

with:

\begin{displaymath}
G(\delta,q) = \int_{\delta_{l}(\delta)}^{\infty}P_{2}(\delta_{l},q=r(1+\delta)^{\frac{1}{3}})~d\delta_{l}
\end{displaymath}

where the dependence on $q$ enters through the coefficients.

Inserting expression (\ref{eqA7}) into expressions (\ref{eq12}) and (\ref{eq21}) in the place of $P(\delta,r)$, 
we may obtain the values of $\overline{\delta}(r)$, $\overline{V}(r)$ (the average corresponding to $\delta$ values 
between $\delta_{min}$ and $\delta_{max}$) as given by our best approximation to the probability distribution for 
$\delta$ at a given r. After integrating by parts we find:

\begin{displaymath}
\overline{\delta}(r) = \int_{\delta_{min}}^{\delta_{max}}G(\delta,r)~d\delta -\delta_{min} -\delta_{max}~G(\delta_{max},r)
\end{displaymath}
\begin{equation}
\overline{V_{r}}(r) = r~\left(\int_{\delta_{min}}^{\delta_{max}}f'(\delta)~G(\delta,r)~d\delta +f(\delta_{min}) 
-f(\delta_{max})~G(\delta_{max},r)\right)  
\label{eqA8}
\end{equation}

where $f'(\delta)$ stands for the derivative of $f(\delta)$ (defined in expression (\ref{eq19})) with respect to $\delta$.

\section{Appendix B}  \label{apendB}

Let $P(\delta_{l},q)$ be the probability distribution for the linear density fluctuation, $\delta_{l}$, at a fixed Lagrangian distance, 
$q$, from the center of an object and let $P(\delta,r)$ be the probability distribution for the actual density fluctuation, $\delta$, 
at a fixed Eulerian distance, $r$, from the center of the same object. We shall show that the following relationship holds:

\begin{equation}
\int_{\delta_{0}}^{\infty}P(\delta,r)~d\delta = \int_{\delta_{l}(\delta_{0})}^{\infty}P(\delta_{l},q=r(1+\delta_{0}) 
^{\frac{1}{3}})~d\delta_{l}                  \label{eqB1}
\end{equation}

from which expressions (\ref{eq9}) and (\ref{eq24}) immediately follow.

If $\delta_{l}$ takes the value $\delta_{l}(\delta_{0})$ at Lagrangian radii $r(1+\delta_{0})^{\frac{1}{3}}$, it is clear that, 
by construction, $\delta$ must take the value $\delta_{0}$ at Eulerian radii, $r$. Now, if at Eulerian radii $r$, $\delta$ takes a 
value , $\delta'$, larger than $\delta_{0}$, its corresponding Lagrangian radii, $q'$, must satisfy:

\begin{displaymath}
q' = r~(1+\delta')^{\frac{1}{3}} > r~(1+\delta_{0})^{\frac{1}{3}} = q
\end{displaymath}

and the value of $\delta_{l}$ at $q'$, $\delta'_{l}=\delta_{l}(\delta')$, must be larger than $\delta_{l}(\delta_{0})$. But, if the 
linear profile is monotonically decreasing the value of the linear density fluctuation at $q$, $\delta_{l}$, must be larger than the 
value at $q'$, $\delta'_{l}$, which is, ex hypothesis, larger than $\delta_{l}(\delta_{0})$. So, whenever $\delta$ is larger 
than $\delta_{0}$ at $r$, $\delta_{l}$ is larger than $\delta_{l}(\delta_{0})$ at $q=r(1+\delta_{0})^{\frac{1}{3}}$, hence, 
(\ref{eqB1}) follows. Note that, if $P(\delta_l,q)$ vanishes for $\delta_l>\delta_{vir}$, the upper limit in the right hand 
side of (\ref{eqB1}) may be set equal to $\delta_{vir}$.


\begin{thebibliography}{40}
 
\bibitem[Avila-Reese et al. 1998]{avila} Avila-Reese, V., Firmani, C. \& Hernandez, X. \ 1998, \apj, 505, 37
\bibitem[Ascasibar et al. 2004]{Yago} Ascasibar, Y., Yepes, G., Gottloeber, S. \& Mueller, V. \ 2004, \mnras, 352, 1109
\bibitem[]{Bajtlik} Bajtlik S., Duncan R.~C., Ostriker J.~P., 1988, \apj, 327, 570
\bibitem[Bardeen et al. 1986]{bardeen} Bardeen, J. M., Bond, J. R., Kaiser, N.
 \& Szalay, A. S. \ 1986, \apj, 304, 15 (BBKS)
\bibitem[Barkana 2004]{Barkana} Barkana R., \ 2004, \mnras, 347, 59
\bibitem[Bernardeau 1994]{Bernie} Bernardeau, F. \ 1994, \apj, 427, 51
\bibitem[Bertschinger 1985]{bert} Bertschinger, E. \ 1985, \apjs, 58,39
\bibitem[Brainerd (2004)]{Brainerd04} Brainerd, T.G., 2004, astro-ph/0409381
\bibitem[Conroy~et al.(2004)]{DEEP} Conroy, C., Newman, J. A., Davis, M., Coil, A. L., Yan, R., Cooper, M. C., 
 Gerke, B. F., Faber, S. M., \& Koo, D. C., 2004, astro-ph/0409305
\bibitem[Fillmore \& Goldreich 1984]{fillmore} Fillmore, J. A. \& Goldreich, P. \ 1984, \apj, 281, 1
\bibitem[Gunn \& Gott 1972]{gunn72} Gunn, J. E. \& Gott, J. R. \ 1972, \apj, 176, 1
\bibitem[Gunn 1977]{gunn77} Gunn J.~E., 1977, \apj, 218, 592
\bibitem[Gurevich \& Zybin 1988]{gura} Gurevich, A. V. \& Zybin, K. P. \ 1988, ZHETF, 94, 3
\bibitem[Guzik \& Seljak (2002)]{GS02} Guzik, J., \& Seljak, U., 2002, \mnras, 335, 311
\bibitem[Hoekstra~et al.(2004)]{Hoekstra} Hoekstra, H., Yee, H. K. C., Gladders, M. D., 2004, \apj, 606, 67
\bibitem[Hiotelis 2001]{griego} Hiotelis, N. \ 2002, \aa, 382, 84
\bibitem[Hoffman \& Shaham 1985]{hoffman} Hoffman Y., Shaham J., 1985, \apj, 297, 16
\bibitem[Jenkins et al. 2001]{jenkins} Jenkins A., Frenk C.~S., White S.~D.~M., Colberg J.~M., Cole S., Evrard A.~E., 
Couchman H.~M.~P., Yoshida N., 2001, \mnras, 321, 372
\bibitem[Kravtsov~et al.(1997)]{Kravtsov97} Kravtsov, A.V., Klypin, A.A.,
 \& Khokhlov, A.M., 1997, \apjs, 111, 73 
\bibitem[Lilje et al. 1986]{lilje}  Lilje, P.B., Yahil, A. \& Jones, B.J.T. \ 1986, \apj, 307, L91
\bibitem[Lokas 2000]{lokas} Lokas E.~L., 2000, \mnras, 311, 423
\bibitem[Lokas \& Hoffman 2000]{lokas01} Lokas, E. L. \& Hoffman, Y. 2000, \apj, 542, L139
\bibitem[Nusser 2001]{nusser} Nusser, A. 2001, \mnras, 325, 1397
\bibitem[Padmanabhan 1996]{padmanaban} Padmanabhan T., 1996, \mnras, 278, L29
\bibitem[Patiri et al. 2004]{Pat} Patiri S., Betancort-Rijo J.~E. \& Prada F. \ 2004, astro-ph/0407513
\bibitem[Prada~et al.(2003)]{Prada2003} 
Prada, F., Vitvitska, M., Klypin, A., Holtzman, J.~A., Schlegel, D.~J., Grebel, E.~K., 
Rix, H.-W., Brinkmann, J., McKay, T.~A., \& Csabai, I., 2003, \apj, 598, 260
\bibitem[Prada et al. 2005]{Paquin} Prada F., Klypin A.~A., Simmoneau E., 
 Betancort-Rijo J., Patiri S.~G., Gottl\"ober S., Sanchez-Conde M.~A., \ 2005 astro-ph/0506432
\bibitem[Ryden \& Gunn 1987]{ryden87} Ryden, B. S. \& Gunn, J. E. \ 1987, \apj, 318, 15
\bibitem[Sheldon~et al.(2004)]{Sheldon} Sheldon, E. S., et al., 2004, \aj, 127, 2544
\bibitem[Sheth and Tormen 2002]{ST02} Sheth, R. K. \& Tormen, G. \ 2002, \mnras, 329, 61
\bibitem[Sikivie et al. 1997]{siki} Sikivie, P., Tkachev, I. I. \& Wang Y. \ 1997,
 \emph{Phys. Rev. D}, 56(4), 1863 .
\bibitem[Smith~et al.(2001)]{Smith01} Smith, D. R., Bernstein, G. M., Fischer, P., \& Jarvis, M., 2001, \apj, 551, 643
\bibitem[Subramanian, Cen \& Ostriker]{subramanian} Subramanian K., Cen R., Ostriker J.~P., 2000, \apj, 538, 528
\bibitem[White \& Zaritsky 1992]{white} White, S. D. M. \& Zaritsky, D. \ 1992, \apj, 394, 1
\bibitem[Zaritsky \& White(1994)]{ZaritskyWhite}Zaritsky, D.~\& White,
S.~D.~M.\ 1994, \apj, 435, 599
\bibitem[Zaritsky et al.(1997)]{Zaritsky97}Zaritsky, D., Smith, R.,
  Frenk, C., \& White, S.~D.~M.\ 1997, \apj, 478, 39 
\end{thebibliography}
\end{document}